\documentclass[10pt,twocolumn]{article} 
\usepackage{latex8}
\usepackage{amsmath,amsthm,amssymb}
\usepackage{proof}
\usepackage{xspace}
\usepackage{times}
\usepackage{graphics} 

\long\def\ignore#1{}

\newtheorem{defn}{Definition}
\newtheorem{lemma}[defn]{Lemma}
\newtheorem{theorem}[defn]{Theorem}

\newtheorem{cor}[defn]{Corollary}

\newcommand{\lvl}{{\rm lvl}}
\newcommand{\supp}{{\rm supp}}
\newcommand{\rulespace}{\vspace{-0.5cm}}

\newcommand{\abs}[1]{\hbox{\sl abs} \; #1}
\newcommand{\tabs}[2]{\hbox{\sl abs} \; #1 \; #2}
\newcommand{\app}[2]{\hbox{\sl app} \; #1 \; #2}
\newcommand{\arr}[2]{\hbox{\sl arr} \; #1 \; #2}
\newcommand{\of}[2]{\hbox{\sl of} \; #1 \; #2}
\newcommand{\member}[2]{\hbox{\sl member} \; #1 \; #2}
\newcommand{\element}[3]{\hbox{\sl element}_{#1} \; #2 \; #3}
\newcommand{\seq}[3]{\hbox{\sl seq}_{#1} \; #2 \; #3}
\newcommand{\prog}[2]{\hbox{\sl prog} \; #1 \; #2}
\newcommand{\cntx}[1]{\hbox{\sl cntx} \; #1}

\newcommand{\subst}[4]{\hbox{\sl subst}_{#1} \; #2 \; #3 \; #4}
\newcommand{\nat}[1]{\hbox{\sl nat} \; #1}
\newcommand{\name}[1]{\hbox{\sl name} \; #1}
\newcommand{\fresh}[2]{\hbox{\sl fresh} \; #1 \; #2}
\newcommand{\tup}[1]{\langle #1\rangle}
\newcommand{\tridot}{\!\Vdash\!} 

\newcommand{\parag}[1]{\smallskip\noindent{\bf #1}\quad}

\newcommand{\FOL   }{FO\lambda}
\newcommand{\N}{{\rm I} \! {\rm N}}
\newcommand{\FOLDN }{\FOL^{\Delta\N}}
\newcommand{\foldnb}{$FO\lambda^{\Delta\nabla}$\xspace}
\newcommand{\logic}{${\cal G}$\xspace}
\newcommand{\LG}{$LG^\omega$\xspace}
\newcommand{\ie}{{\em i.e.}} 
\newcommand{\eg}{{\em e.g.}} 
\newcommand{\etc}{{\em etc}} %
\newcommand{\lp}{$\lambda$Prolog\xspace}

\newcommand{\defL}{\hbox{\sl def}\mathcal{L}}
\newcommand{\defR}{\hbox{\sl def}\mathcal{R}}
\newcommand{\cL}{\hbox{\sl c}\mathcal{L}}
\newcommand{\natL}{\hbox{\sl nat}\mathcal{L}}
\newcommand{\natR}{\hbox{\sl nat}\mathcal{R}}
\newcommand{\cut}{\hbox{\sl cut}}

\title{Combining generic judgments with recursive definitions}
\author{
Andrew Gacek\\
Department of CS\&E\\
University of Minnesota\\
\and
Dale Miller\\
INRIA Saclay - \^Ile-de-France\\
\& LIX/\'Ecole polytechnique\\
\and 
Gopalan Nadathur\\
Department of CS\&E\\
University of Minnesota\\
}

\begin{document}
\maketitle
\thispagestyle{empty}


\begin{abstract}
Many semantical aspects of programming languages, such as their
operational semantics and their type assignment calculi, are specified
by describing appropriate proof systems. 
Recent research has identified two proof-theoretic features that
allow direct, logic-based reasoning about such descriptions:
the treatment of atomic judgments as fixed points 
(recursive definitions) and an encoding of 
binding constructs via generic judgments.
However, the logics encompassing these two features have thus far
treated them orthogonally: that is, 
they do not provide the ability to define
object-logic properties that themselves depend on an intrinsic
treatment of binding.
We propose a new and simple integration of these features within an
intuitionistic logic enhanced with induction over natural numbers and
we show that the resulting logic is consistent.
The pivotal benefit of the integration is that it allows recursive
definitions to not just encode simple, traditional forms of {\em
  atomic judgments}  but also to capture generic properties pertaining
to such judgments. 
The usefulness of this logic is illustrated by showing how it can
provide elegant
treatments of object-logic contexts that
appear in proofs involving typing calculi and of arbitrarily cascading
substitutions that play a role in reducibility arguments.
\end{abstract}




\Section{Introduction}
\label{sec:introduction}

An important approach to specifying and reasoning about computations involves
{\em proof theory} and {\em proof search}.  We
discuss below three kinds of judgments about computational
systems that one might want to capture and the proof theoretic
techniques that have been used to capture them. We divide this
discussion into two parts: the first part deals with judgments over
{\em   algebraic terms} and the second with judgments over {\em
  terms-with-binders}. We then exploit this overview to describe the new
features of the logic we are presenting in this paper.

\SubSection{Judgments involving algebraic terms}

We overview features of proof theory that support recursive
definitions about first-order
(algebraic) terms and, using CCS as an example, we illustrate the
judgments about computations that can be encoded through such 
definitions.

\parag{(1) Logic programming, may behavior}
Logic programming languages allow for a natural
specification and animation of operational semantics and typing
judgments: this observation goes back to at least the Centaur project
and its animation of Typol specifications using Prolog
\cite{borras88}.  For example, Horn clauses provide a simple and
immediate encoding of CCS labeled transition systems and 
unification and backtracking provide a means for
exploring what is {\em reachable} from a given process. 
Traditional logic programming is, however, limited to {\em may} behavior
judgments: using it, we cannot prove that a
given CCS process $P$ cannot make a transition and, since this
negative property
is logically equivalent to proving that $P$ is bisimilar to $0$ (the
null process), such systems cannot capture bisimulation.

\parag{(2) Model checking, must behavior}
Proof theoretic techniques for {\em must} behaviors (such as
bisimulation and many model checking problems) have been developed in
the early 1990's \cite{girard92mail,schroeder-Heister93lics} and
further extended later \cite{mcdowell00tcs}. Since these techniques work
by unfolding computations until termination, they are applicable to
{\em recursive definitions} that are {\em noetherian}. As an example, 
bisimulation for finite CCS can be given an immediate and declarative
specification \cite{mcdowell03tcs}.

\parag{(3) Theorem proving, infinite behavior}
Reasoning about all members of a domain or about
possibly infinite executions requires induction and coinduction.
Incorporating induction in proof theory goes back to Gentzen.  The
work in \cite{mcdowell00tcs,momigliano03types,tiu04phd} provides
induction and coinduction rules associated with the above-mentioned
recursive definitions.  In such a setting, one can prove, for example,
that (strong) bisimulation in CCS is a congruence.

\SubSection{Judgments involving bindings}

The proof theoretic treatment of binding in terms has echoed the three
stages of development described above. We switch from CCS to the
$\pi$-calculus to illustrate the different kinds of judgments that
these support.

\parag{(1) Logic programming, $\lambda$-tree syntax} Higher-order
generalizations of logic programming, such as {\em higher-order
hereditary Harrop formulas} \cite{miller91apal} and the dependently
typed LF \cite{harper93jacm}, adequately capture may behavior for
terms containing bindings.  In particular, the presence of
hypothetical and universal judgments supports the $\lambda$-tree
syntax \cite{miller00cl} approach to higher-order abstract syntax
\cite{pfenning88pldi}.  The logic programming languages \lp\
\cite{nadathur88iclp} and Twelf \cite{pfenning99cade} support such
syntax representations and provide simple specification of, for
example, reachability in the $\pi$-calculus.

\parag{(2) Model checking, $\nabla$-quantification}
While the notions of universal 
quantification and {\em generic judgment} are often conflated, a
satisfactory treatment of must behavior requires splitting apart these
concepts.  The $\nabla$-quantifier \cite{miller05tocl} was introduced 
to encode generic judgments directly.  To
illustrate the issues here, consider the formula $\forall
w.\neg (\lambda x.x = \lambda x.w).$ If we think of $\lambda$-terms as
denoting abstracted syntax (terms modulo $\alpha$-conversion), this
formula should be provable (variable capture is not allowed in logically
sound substitution).  If we think of $\lambda$-terms as describing
functions, then the equation $\lambda y.t = \lambda y.s$ is equivalent
to $\forall y. t = s$.  But then our example formula is equivalent to 
$\forall w.\neg \forall x.x = w$, which should not be provable since it
is not true in a model with a single element domain. 
To think of $\lambda$-terms syntactically instead, we treat $\lambda
y.t = \lambda y.s$ as equivalent to $\nabla y. t = s$.  In this case,
our example formula is equivalent to $\forall w.\neg \nabla x.x = w$,
which is provable 
\cite{miller05tocl}. Using this quantifier, the $\pi$-calculus process $(\nu
x).[x=w].\bar w x$ can be encoded such that it is provably 
bisimilar to $0$.  Bedwyr \cite{baelde07cade} is a model checker that
treats such generic judgments.

\parag{(3) Theorem proving,~ \LG}
When there is only finite behavior, logics for recursive definitions
do not need the cut or initial rules, and, consequently, they
do not need to answer the question ``When are two generic judgments
equal?''  On the other hand, induction and coinduction do need an
answer to this question: \eg, when doing induction over natural
numbers, one must be able to recognize that the case for $i+1$ has
been reduced to the case for $i$.  The \LG proof system 
\cite{tiu06lfmtp}
provides a natural setting for answering this question. 
Using \LG encodings, one
can prove that (open) bisimulation is a $\pi$-calculus congruence.

\SubSection{Allowing definitions of generic judgments}

In the developments discussed above, recursive definitions are
permitted only for {\em atomic} judgments.  In many syntax analysis
problems, binding constructs are treated by building up a local
context that attributes properties to the objects they bind. In
reasoning about such analyses, it is often necessary to be able to
associate relevant {\em generic} properties with atomic judgments.
For example, a typical type assignment calculus for $\lambda$-terms
treats abstractions by adding assumptions about the type of the bound
variables to the context of the typing judgment.  To model such a
context, we might use a predicate {\sl cntx} that encodes the
assignment of types to abstracted variables. Thus, an atomic judgment
of the form $\cntx [\tup{x_1, t_1},\ldots,\tup{x_n, t_n}]$ would
denote the assignment of types $t_1,\ldots,t_n$ to the variables
$x_1,\ldots,x_n$ and can be used as a hypothesis in the course of
determining the type of a term.  Now, certain ``generic'' properties
hold implicitly of the contexts that are constructed: for example,
these assign types only to bound variables and have at most one
assignment for each of them. Such properties are
not actually used in encoding the rules for type inference but they do
have to be made explicit if we want to prove properties, such as the
determinacy of type assignment, about the calculus that is
encoded. Recursive definitions provide a means for formalizing
properties that are needed in these kinds of reasoning
tasks. Unfortunately, these definitions are not strong enough in their
present form to allow for the convenient statement of generic
properties ranging over atomic judgments.  

These issues
surrounding the specification of contexts
are actually endemic to 
reasoning about many different kinds of specifications that utilize
$\lambda$-tree syntax. We provide an elegant treatment of it here by
extending recursive definitions to apply not only to 
atomic but also to generic judgments. Using this device, we will,
for instance, be able to define a property of the form 
\[
\nabla x_1\cdots\nabla x_n.\ \cntx [\tup{x_1, t_1},\ldots,\tup{x_n,
    t_n}].
\]
By stating the property in this way, we ensure that \hbox{\sl cntx}
assigns types only to variables and at most one to each. Now, this
property can be used in an inductive proof, provided it can be
verified that the contexts that are built up during type analysis
recursively satisfy the definition.  We present rules that support
this style of argument.

\SubSection{An outline of the paper}

Section~\ref{sec:logic} describes the logic \logic that 
allows for the extended form of definitions and
Section~\ref{sec:meta-theory} establishes its consistency. 
The extension has significant consequences for writing and reasoning
about logical specifications. We provide a
hint of this through a few examples in Section
~\ref{sec:examples}; as discussed later, many other
applications such as solutions to the POPLmark challenge problems
\cite{aydemir05tphols}, cut-elimination for sequent calculi,
and an encoding of Tait's logical relations
based proof of 
normalization for the simply typed $\lambda$-calculus
\cite{tait67jsl} have been successfully developed 
using the Abella system that implements \logic. We conclude the
paper with a comparison to related work and an indication of future
directions. 



\Section{A logic with generalized definitions}
\label{sec:logic}

The logic \logic is obtained by extending an intuitionistic and
predicative subset of Church's Simple Theory of Types with fixed point
definitions, natural number induction, and a new quantifier for
encoding generic judgments. Its main components are 
elaborated in the subsections below.  It is possible to develop a
classical variant of \logic as well: we do not follow that path but
just comment that moving from intuitionistic to classical logic can
have interesting impacts on specifications.  For example, the
intuitionistic reading of the specification of bisimulation for the
$\pi$-calculus yields {\em open bisimulation} while the classical reading of
the same specification yields {\em late bisimulation} \cite{tiu04fguc}.

\SubSection{The basic syntax}

Following Church \cite{church40}, terms are constructed
using abstraction and application from constants and (bound)
variables. All terms are typed using a monomorphic typing system;
these types also constrain the set of well-formed expressions in the
expected way.  The provability relation concerns well-formed terms of
the distinguished type $o$ that are also called formulas. Logic is
introduced  
by including special constants representing the propositional
connectives $\top$, $\bot$, $\land$, $\lor$, $\supset$ and, for every
type $\tau$ that does not contain $o$, the constants $\forall_\tau$
and $\exists_\tau$ of type $(\tau \rightarrow o) \rightarrow o$.  The
binary propositional connectives are written as usual in infix form
and the expressions $\forall_\tau x. B$ and $\exists_\tau x. B$
abbreviate the formulas $\forall_\tau \lambda x.B$ and $\exists_\tau
\lambda x.B$, respectively.  Type subscripts will be omitted
from quantified formulas when they can be inferred from the context or
are not important to the discussion. We also use a shorthand for
iterated quantification: if ${\cal Q}$ is a quantifier, the expression
${\cal Q}x_1,\ldots,x_n.P$ will abbreviate ${\cal
  Q}x_1\ldots{\cal Q}x_n.P$.

The usual inference rules for the universal quantifier can be seen as
equating it to the conjunction of all of its instances: that is, this
quantifier is treated extensionally.  
There are a number of situations \cite{miller05tocl} where one wishes
to have a generic treatment of a statement like ``$B(x)$ holds for all
$x$'': in these situations, the {\em form} of the argument is
important and not the argument's behavior on all its possible
instances.  
To encode such
generic judgments, we use the $\nabla$-quantifier (nabla)
\cite{miller05tocl}. Syntactically, this quantifier corresponds to
including a constant $\nabla_\tau$ of type $(\tau \rightarrow o)
\rightarrow o$ for each type $\tau$ (not containing $o$).  As with the
other quantifiers, $\nabla_\tau x. B$ abbreviates $\nabla_\tau \lambda
x. B$ and the type subscripts are often suppressed for readability.

\SubSection{Generic judgments and $\nabla$-quantification}

Sequents in intuitionistic logic are usually written as
\[\Sigma : B_1, \ldots, B_n \vdash B_0 \qquad (n\ge0)
\]
where $\Sigma$ is the ``global signature'' for the sequent: in
particular, it contains the eigenvariables of the sequent proof.
We shall think of $\Sigma$ in this prefix position as being a binding
operator for each variable it contains.  The \foldnb logic 
\cite{miller05tocl} introduced ``local signatures'' for each formula
in the sequent: that is, sequents are written instead as
\[\Sigma : \sigma_1\triangleright B_1, \ldots,
           \sigma_n\triangleright B_n \vdash
           \sigma_0\triangleright B_0,
\]
where each $\sigma_0, \ldots, \sigma_n$ is a list of variables that are
bound locally in the formula adjacent to it.  Such local signatures
within proofs reflect bindings in formulas using the
$\nabla$-quantifier: in particular, the judgment and formula 
\[
  x_1,\ldots,x_n\triangleright B
   \hbox{\quad and\quad}
  \nabla x_1\cdots\nabla x_n.B 
  \qquad (n\ge0)
\]
have the same proof-theoretic force.

The \foldnb logic \cite{miller05tocl} (and its partial
implementation in the Bedwyr logic programming/model checking system
\cite{baelde07cade}) eschewed atomic formulas for explicit fixed point
(recursive) definitions, along with inference rules to unfold them.
In such a system, both the cut-rule and the initial rule can be
eliminated and checking the equality of two generic judgments is not
necessary.  As we have already mentioned, when one is proving more
ambitious theorems involving induction and coinduction, equality of
generic judgments becomes important.

\begin{figure*}[t]
\[
\begin{array}{ccc}
\infer[id_\pi]{\Sigma : \Gamma, B \vdash B'}{\pi.B = \pi'.B'} &
\infer[\cut]{\Sigma : \Gamma, \Delta \vdash C}
            {\Sigma : \Gamma \vdash B & \Sigma : B, \Delta \vdash C} &
\infer[\cL]{\Sigma : \Gamma, B \vdash C}
           {\Sigma : \Gamma, B, B \vdash C}
\\
\noalign{\smallskip}
\infer[\bot\mathcal{L}]{\Sigma :\Gamma,\bot \vdash C}{} &
\infer[\lor\mathcal{L}]{\Sigma :\Gamma,B\lor D \vdash C}
                       {\Sigma :\Gamma,B\vdash C & \Sigma:\Gamma,D\vdash C} &
\infer[\lor\mathcal{R},i\in\{1,2\}]{\Sigma : \Gamma \vdash B_1 \lor B_2}
                                   {\Sigma : \Gamma \vdash B_i}
\\
\noalign{\smallskip}
\infer[\top\mathcal{R}]{\Sigma : \Gamma \vdash \top}{} &
\infer[\land\mathcal{L},i\in\{1,2\}]{\Sigma : \Gamma, B_1 \land B_2
  \vdash C}{\Sigma : \Gamma, B_i \vdash C} &
\infer[\land\mathcal{R}]{\Sigma : \Gamma \vdash B \land C}{\Sigma :
  \Gamma \vdash B & \Sigma : \Gamma \vdash C}
\end{array}
\]\smallskip\[
\begin{array}{cc}
\infer[\supset\mathcal{L}]{\Sigma : \Gamma, B \supset D \vdash C}
      {\Sigma : \Gamma \vdash B & \Sigma : \Gamma, D \vdash C} &
\infer[\supset\mathcal{R}]{\Sigma : \Gamma \vdash B \supset C}
      {\Sigma : \Gamma, B \vdash C}
\\
\noalign{\smallskip}
\infer[\forall\mathcal{L}]{\Sigma : \Gamma, \forall_\tau x.B \vdash C}
      {\Sigma, \mathcal{K}, \mathcal{C} \vdash t : \tau &
       \Sigma : \Gamma, B[t/x] \vdash C} &
\infer[\forall\mathcal{R},h\notin\Sigma,\supp(B)=\{\vec{c}\}]
      {\Sigma : \Gamma \vdash \forall x.B}
      {\Sigma, h : \Gamma \vdash B[h\ \vec{c}/x]}
\\
\noalign{\smallskip}
\infer[\nabla\mathcal{L},a\notin \supp(B)]
      {\Sigma : \Gamma, \nabla x. B \vdash C}
      {\Sigma : \Gamma, B[a/x] \vdash C} &
\infer[\nabla\mathcal{R},a\notin \supp(B)]
      {\Sigma : \Gamma \vdash \nabla x.B}
      {\Sigma : \Gamma \vdash B[a/x]}
\\
\noalign{\smallskip}
\infer[\exists\mathcal{L},h\notin\Sigma,\supp(B)=\{\vec{c}\}]
      {\Sigma : \Gamma, \exists x. B \vdash C}
      {\Sigma, h : \Gamma, B[h\; \vec{c}/x] \vdash C} &
\infer[\exists\mathcal{R}]{\Sigma : \Gamma \vdash \exists_\tau x.B}
      {\Sigma, \mathcal{K}, \mathcal{C} \vdash t:\tau &
       \Sigma : \Gamma \vdash B[t/x]}
\end{array}
\]
\caption{The core rules of \logic}
\label{fig:core-rules}
\end{figure*}

\SubSection{\LG and structural rules for $\nabla$-quantification}
\label{sec:lg}

There are two 
equations for $\nabla$ that we seem forced to
include when we 
consider proofs by induction. In a sense, these
equations play the role of structural rules
for the local, generic context.  Written at the level of formulas,
they are the $\nabla${\em -exchange rule} $\nabla x\nabla y. F =
\nabla y\nabla x. F$ and the $\nabla${\em -strengthening rule} $\nabla
x. F = F$, provided $x$ is not free in $F$.  The 
\LG proof system of Tiu \cite{tiu06lfmtp} is essentially \foldnb
extended with these two structural rules for $\nabla$. 

The move from the weaker \foldnb to the stronger \LG logic has at
least two important additional consequences.

First, the strengthening rule implies that every type at which one is
willing to use $\nabla$-quantification is not only non-empty but
contains an unbounded number of members.  For example, the formulas
$\exists_\tau x.\top$ is always provable, even if there are no closed
terms of type $\tau$ because this formula is equivalent to 
$\nabla_\tau y\exists_\tau x.\top$ which is provable, as will be clear
from the proof system given in Figure~\ref{fig:core-rules}.
Similarly, for any given $n\geq 1$, the following formula 
is provable
\[
\exists x_1\ldots\exists x_n 
 [\bigwedge_{1\leq i,j\leq n, i\not= j} x_i \not= x_j].
\]

Second, the validity of the strengthening and exchange rules mean that
all local contexts can be made equal. As a result, the local binding
can now be considered as an (implicit) global binder.  In such a
setting, the collection of globally $\nabla$-bound variables can be
replaced with {\it nominal constants}. Of course, in light of the
exchange rule, we must consider atomic judgments as being identical if
they differ by only permutations of such constants.  

We shall follow the \LG approach to treating $\nabla$.
Thus, for every type we assume an infinite collection
of nominal 
constants.  The collection of all nominal constants is denoted by
$\mathcal{C}$; these constants are to be distinguished from the
collection of usual, non-nominal constants that we denote by 
$\mathcal{K}$. 
We define the {\it support} of a term (or 
formula), written $\supp(t)$, as the set of nominal constants 
appearing in it. 
A permutation of nominal constants is a bijection $\pi$ from
$\mathcal{C}$ to $\mathcal{C}$ such that $\{ x\ |\ \pi(x) \neq x\}$ is
finite and $\pi$ preserves types. Permutations will be extended to
terms (and formulas), written $\pi . t$, as follows:
\[
\begin{array}{l@{\quad }l}
\pi.a = \pi(a), \mbox{ if $a \in \mathcal{C}$} &
\pi.c = c, \mbox{ if $c\notin \mathcal{C}$ is atomic} \\
\pi.(\lambda x.M) = \lambda x.(\pi.M) &
\pi.(M\; N) = (\pi.M)\; (\pi.N) 
\end{array}
\]

The core fragment of \logic is presented in Figure
\ref{fig:core-rules}. Sequents in this logic have the form $\Sigma :
\Gamma \vdash C$ where $\Gamma$ is a multiset and the signature
$\Sigma$ contains all the free variables of $\Gamma$ and $C$. In the
$\nabla\mathcal{L}$ and $\nabla\mathcal{R}$ rules, $a$ denotes a
nominal constant of an appropriate type. In the $\exists\mathcal{L}$
and $\forall\mathcal{R}$ rule we use raising \cite{miller92jsc} to
encode the dependency of the quantified variable on the support of
$B$; the expression $(h\ \vec{c})$ used in these two rules denotes the
(curried) application of $h$ to the constants appearing in the
sequence $\vec{c}$. The
$\forall\mathcal{L}$ and $\exists\mathcal{R}$ rules make use of
judgments of the form $\Sigma, \mathcal{K}, \mathcal{C} \vdash t :
\tau$. These judgments enforce the requirement that the expression
$t$ instantiating the quantifier in the rule is a well-formed term of
type $\tau$ constructed from the variables in $\Sigma$ and the
constants in ${\cal K} \cup {\cal C}$. Notice that in contrast the
$\forall\mathcal{R}$ and $\exists\mathcal{L}$ rules seem to allow for
a dependency on only a restricted set of nominal constants. However,
this asymmetry is not significant: the dependency expressed through
raising in the latter rules can be extended to any
number of nominal constants that are not in the relevant support set
without affecting the provability of sequents.

\SubSection{Recursive definitions}

The structure of definitions in \logic is, in a sense, its
distinguishing characteristic. To motivate their form and also to
understand their expressiveness, we consider first the definitions
that are permitted in $LG^\omega$. In that setting, a
definitional clause has the form  $\forall \vec{x}. H
\triangleq B$ where $H$ is an atomic formula all of whose free
variables are contained in $\vec{x}$ and $B$ is an arbitrary formula
all of whose free variables must also be free in $H$. In a clause of
this sort, $H$ is called the {\em head} and $B$ is called the {\em
  body} and a (possibly infinite) collection of clauses constitutes a
definition. 
Now, there are two properties of such definitional clauses that should
be noted. First, $H$ and $B$ are restricted to not contain occurrences
of nominal constants. Second, the
interpretation of such a clause permits the variables in $\vec{x}$ to
be instantiated with terms containing {\em any} nominal constant;
intuitively, the quantificational structure at the head of the
definition has a $\nabla \forall$ form, with the (implicit) $\nabla$ 
quantification being over arbitrary sequences of nominal
constants. These two properties actually limit the power of
definitions: (subparts of) terms satisfying the relations they
identify cannot be forced to be nominal constants and, similarly,
specific (sub)terms cannot be stipulated to be independent of such
constants.  
  
These shortcomings are addressed in \logic by allowing definitional
clauses to take the form 
$\forall \vec{x} .(\nabla \vec{z} . H) \triangleq B$
where all the free variables in $\nabla \vec{z} . H$ must appear in
$\vec{x}$ and all the free variables in $B$ must also be free in
$\nabla \vec{z} . H$.
The intended interpretation of the $\nabla$ quantification over $H$ is
that particular terms appearing in the relation being defined must be 
identified as nominal constants although specific names may still not
be assigned to these constants. Moreover, the location of this
quantifier changes the prefix over the head from a
$\nabla \forall$ form to the more general 
$\nabla \forall \nabla$ form.  Concretely, the explicit $\nabla$
quantification over $\vec{z}$ forces the instantiations for the 
externally $\forall$ quantified variables $\vec{x}$ to be independent
of the nominal constants used for $\vec{z}$.  

One illustration of the definitions permitted in \logic is provided by the
following clause:
\begin{tabbing}
\qquad\=\kill
\>$(\nabla n . \name n) \triangleq \top$.
\end{tabbing}
An atomic predicate $\name N$ would satisfy this clause provided that
it can be matched with its head. For this to be possible, $N$
must be a nominal constant. Thus, {\sl name} is a predicate that
recognizes such constants. As another example, consider the clause
\begin{tabbing}
\qquad\=\kill
\>$\forall E . (\nabla x . \fresh x E) \triangleq \top$.
\end{tabbing}
In this case the atomic formula $\fresh N T$ will satisfy the clause
just in case $N$ is a nominal constant and $T$ is a term that does not
contain this constant (the impossibility of variable capture ensures
this constraint).  Thus, this clause expresses the property of a
name being ``fresh'' to a given term. Further illustrations of the new
form of definitions and their use in reasoning tasks are
considered in Section~\ref{sec:examples}.

Definitions impact the logical system through introduction
rules for atomic judgments. Formalizing these rules involves the use
of substitutions. A {\it substitution} $\theta$ is a type-preserving
mapping (whose application is written in postfix notation) from
variables to terms, such 
that the set $\{x\ |\ x\theta \neq x\}$ is finite. Although
a substitution is extended to a mapping from terms to terms, formulas
to formulas, \etc, when we refer to its {\it domain} and {\it range},
we mean these sets for this most basic
function. A substitution is extended to a function from terms to terms
in the usual fashion.
If $\Gamma$ is a multiset 
of formulas then $\Gamma\theta$ is the multiset $\{J\theta\ |\ J \in
\Gamma\}$. If $\Sigma$ is a signature then $\Sigma\theta$ is the
signature that results from removing from $\Sigma$ the variables in
the domain of $\theta$ and adding the variables that are free in the
range of $\theta$.

To support the desired interpretation of a definitional clause, 
when matching the head of $\forall \vec{x} . (\nabla \vec{z} . H)
\triangleq B$ with an atomic judgment, we must permit the
instantiations for $\vec{x}$ to contain the nominal constants
appearing in that judgment. Likewise, we must consider instantiations
for the eigenvariables appearing in the judgment that possibly
contain the nominal constants chosen for $\vec{z}$. Both possibilities
can be realized via raising. Given a clause $\forall
x_1,\ldots,x_n . (\nabla \vec{z} . H) \triangleq B$, we define a 
version of it raised over the sequence of nominal constants $\vec{a}$
and away from a signature $\Sigma$ as 
\begin{multline*}
\forall \vec{h} . (\nabla \vec{z} . H[h_1\; \vec{a}/x_1, \ldots,
h_n\; \vec{a}/x_n]) \triangleq \\
B[h_1\; \vec{a}/x_1, \ldots, h_n\; \vec{a}/x_n],
\end{multline*}
where $h_1,\ldots,h_n$ are distinct variables of suitable type that do
not appear in $\Sigma$. 
Given the sequent $\Sigma : \Gamma \vdash C$ and a sequence of nominal
constants $\vec{c}$ none of which appear in the support of $\Gamma$ or
$C$, let $\sigma$ be any substitution of the form
\begin{tabbing}
\qquad\quad\=$\{h'\; \vec{c}/h\ |\ $\=\kill
\>$\{h'\; \vec{c}/h\ |\ h \in \Sigma\ \mbox{and}\ h'\ \mbox{is a
variable of}$\\
\>\> $\mbox{suitable type that is not in}\ \Sigma\}$.
\end{tabbing}
Then the sequent $\Sigma\sigma : \Gamma\sigma \vdash C\sigma$
constitutes a version of $\Sigma : \Gamma \vdash C$ raised over
$\vec{c}$. 

\begin{figure}[t]
\[
\infer[\kern-1pt\defL]{\Sigma : A, \Gamma \vdash C}{\{\Sigma'\theta :
  (\pi.B')\theta, \Gamma'\theta \vdash C'\theta\}}
\ \ 
\infer[\kern-1pt\defR]{\Sigma : \Gamma \vdash A}{\Sigma' : \Gamma'
  \vdash (\pi . B')\theta}
\]
\caption{Rules for definitions}
\label{fig:def}
\end{figure}

The introduction rules based on definitions are presented in
Figure~\ref{fig:def}. The $\defL$ rule has a set of premises
that is generated by considering each definitional clause of the form
$\forall \vec{x}.(\nabla \vec{z}. H) \triangleq B$ in the following
fashion. Assuming that $\vec{z} = z_1,\ldots,z_n$, let $\vec{c}=
c_1,\ldots,c_n$ be a sequence of distinct nominal constants 
none of which appear in the
support of  $\Gamma$, $A$ or $C$ and let $\Sigma' : A', \Gamma' \vdash
C'$  denote a version of the lower sequent raised over 
$\vec{c}$. Further, let $H'$ and $B'$ be obtained by taking the head and
body of a version of the clause being considered raised over a listing
$\vec{a}$ of the constants in the support of $A$ and away from
$\Sigma'$ and applying the substitution 
$[c_1/z_1,\ldots,c_n/z_n]$ to them. Then the set of premises arising
from this clause are obtained by considering all permutations $\pi$ of
$\vec{a}\vec{c}$ and all substitutions $\theta$ such that
$(\pi. H')\theta = A'\theta$, with the proviso that the range of
$\theta$ may not contain any nominal constants. 

The $\defR$ rule has exactly one premise that is obtained
by using any one definitional clause. The formulas $B'$ and $H'$ are
generated from 
this clause as in the $\defL$ case, but $\pi$ is now taken to
be any one permutation of $\vec{a}\vec{c}$  and $\theta$ is taken
to be any one substitution such that $(\pi . H')\theta = A'$, again
with the proviso that the range of $\theta$ may not contain any
nominal constants. 

In summary, the definition rules are based on raising the
sequent over the nominal constants picked for the $\nabla$ variables
from the definition, raising the definition over nominal constants
from the sequent, and then unifying the chosen atomic judgment and
the head of the definition under various permutations of the
nominal constants. 
\ignore{
Occasionally, 
raising the sequent over new nominal constants will be superfluous
since the permutation $\pi$ might not map a $\nabla$ variable from the
definition head to one of these new nominal constants. Instead, the
new nominal constant over which the sequent was raised may match with
a nominal constant from $\vec{a}$ over which the clause was raised.
The net effect of the raising in this case is that a new nominal
constant is introduced but eventually not used in any significant
way. In Section  \ref{sec:meta-theory} we state a lemma which say that
such nominal constants can be ignored or ``un-raised.''
}
As it is stated, the set of premises in the $\defL$ rule arising from
any one definitional clause is potentially infinite because of the
need to consider every unifying substitution. It is possible
to restrict these substitutions instead to the members of a complete
set of unifiers. In the situations where there is a single most
general unifier, as is the case when we are dealing with
the higher-order pattern fragment \cite{miller91jlc}, the
number of premises arising from each definition clause is bounded by
the number of permutations. In practice, this number can be quite
small as illustrated in Section~\ref{sec:examples}.

Two restrictions must be placed on definitional clauses to ensure
consistency of the logic. The first is that no nominal constants may
appear in such a clause; this requirement also enforces an
equivariance property for definitions. The second is 
that such clauses must be {\it stratified} so as to guarantee the
existence of fixed points. To do this we associate with each predicate
$p$ a natural number $\lvl(p)$, the {\it level} of $p$. The notion is
generalized to formulas as follows.
\begin{defn}
Given a formula B, its  level $\lvl(B)$ is defined as follows:
\begin{enumerate}
\item $\lvl(p\ \bar{t}) = \lvl(p)$
\item $\lvl(\bot) = \lvl(\top) = 0$
\item $\lvl(B\land C) = \lvl(B\lor C) = \max(\lvl(B),\lvl(C))$
\item $\lvl(B\supset C) = \max(\lvl(B)+1,\lvl(C))$
\item $\lvl(\forall x.B) = \lvl(\nabla x.B) = \lvl(\exists x.B) = \lvl(B)$
\end{enumerate}
\end{defn}
For every definitional clause $\forall \vec{x} . (\nabla \vec{z}. H)
\triangleq B$, we require $\lvl(B) \leq \lvl(H)$. This stratification
condition ensures that a definition cannot depend negatively on
itself. More precise stratification conditions which allow such
dependency in a controlled fashion are possible, but we choose this
condition for simplicity.
See \cite{mcdowell00tcs,tiu06lfmtp} for a description of why these
properties lead to consistency.

\SubSection{Induction over natural numbers}

\begin{figure}[t]
\begin{equation*}
\infer[\natL]{\Sigma : \Gamma, \nat N \vdash C}
             {\vdash I\; z \qquad
              x : I\; x \vdash I\; (s\; x)\qquad 
              \Sigma : \Gamma, I\; N \vdash C}
\end{equation*}
\rulespace
\begin{align*}
\infer[\natR]{\Sigma : \Gamma \vdash \nat z}{} &&
\infer[\natR]{\Sigma : \Gamma \vdash \nat (s\; N)}{\Sigma :
  \Gamma \vdash \nat N}
\end{align*}
\caption{Rules for natural number induction}
\label{fig:induction}
\end{figure}

The final component of \logic is an encoding of natural numbers and
rules for carrying out induction over these numbers. This form of
induction is useful in reasoning about specifications of computations
because it allows us to induct on the height of object-logic proof
trees that encode the lengths of 
computations. Specifically, we introduce the type $nt$ and
corresponding constructors $z : nt$ and $s : nt \to nt$. Use of
induction is controlled by the distinguished predicate $\hbox{\sl nat}
: nt \to o$. The rules for this predicate are presented in Figure
\ref{fig:induction}. The rule $\natL$ is actually a rule
schema, parameterized by the induction invariant $I$.  Providing
induction over only natural numbers is mostly a matter of
convenience in studying the meta-theory of \logic.  Extending induction
to other algebraic datatypes \cite{momigliano03types,tiu04phd} should
have little impact on the meta-theory of \logic, although it
would clearly be a useful extension for any system implementing \logic
(such as Abella \cite{gacek08abella}).



\Section{Cut-elimination and consistency for \logic}
\label{sec:meta-theory}

The consistency of \logic is an immediate consequence of the
cut-elimination result for this logic. Cut-elimination is
proved for $LG^\omega$ \cite{tiu08lgext} by a generalization of the
approach used for $\FOLDN$ \cite{mcdowell00tcs} that is itself based
on a technique introduced by Tait \cite{tait67jsl} and refined by
Martin-L\"{o}f \cite{martin-lof71sls}. The main aspect of
this generalization is 
recognizing and utilizing the fact that certain transformations of
sequents
preserve provability and also do not increase (minimum) proof
height. The particular transformations that are considered in the case
of \LG have to do with weakening of hypotheses, permutations
of nominal constants, and substitutions for eigenvariables. 
We can use this framework to show that cut can be eliminated from
\logic by adding one more transformation to this collection. This
transformation pertains to the raising of sequents that is needed in
the introduction rules based on the extended form of definitional
clauses. We motivate this transformation by sketching the structure of
the argument as it concerns the use of such clauses below.

The critical part of the cut-elimination argument is the reduction of
what are called the essential cases of the use of the \cut\ rule, \ie,
the situations where the last rule in the derivation is a \cut\ and the last
rules in the derivations of its premises introduce the cut
formula. Now, the only rules of \logic that are different from those of \LG
are $\defL$ and $\defR$. Thus, we have to
consider a different argument only when these rules are the last ones
used in the premise derivations in an essential case of a \cut. In this
case, the overall derivation has the form 
{\small
\begin{equation*}
\infer[\cut]{\Sigma : \Gamma, \Delta \vdash C}{
\infer[\kern-3pt\defR]{\Sigma : \Gamma \vdash A}{\deduce{\Sigma' : \Gamma'
  \vdash (\pi . B')\theta}{\Pi_1}} &\kern-3pt
\infer[\kern-3pt\defL]{\Sigma : A, \Delta \vdash C}
  {\left\{\raisebox{-1.5ex}
         {\deduce{\Sigma''\rho : (\pi'.B'')\rho, \Delta''\rho
             \vdash C''\rho} 
                 {\Pi_2^{\rho,\pi',B''}}}
   \right\}}}
\end{equation*}}%
\noindent where $\Pi_1$ and $\Pi_2^{\rho,\pi',B''}$ represent
derivations of the relevant sequents.
Let $\Sigma' : \Gamma' \vdash A'$ be the raised version of $\Sigma : 
\Gamma \vdash A$ and let $H'$ and $B'$ be the head and body of the
version of the definitional clause raised over $\supp(A)$ and away from
$\Sigma'$ used in the $\defR$ rule.  
From the definition of this rule, we know that $\theta$ is
substitution such that $(\pi. H')\theta = A'$. Let $\theta'$ be the
restriction of $\theta$ to the free variables of $H'$. Clearly $(\pi. H')
\theta = (\pi. H') \theta'$ and $(\pi. B')\theta = (\pi.
B')\theta'$. Further, since the free variables of $H'$ are 
distinct from the variables in 
$\Sigma'$, $\theta'$ has no effect on $\Sigma'$, $\Delta'$, $C'$, or
$A'$. Thus, 
it must be the case that $(\pi. H')\theta' = A'\theta'$. From this it
follows that 
\begin{equation*}
\deduce{\Sigma' : (\pi.B')\theta', \Delta'
             \vdash C'} 
                 {\Pi_2^{\theta',\pi,B'}}
\end{equation*}
is included in the set of derivations above the lower sequent of the
$\defL$ rule. We can therefore reduce the \cut\ in question to the following:
\begin{equation*}
\infer{\Sigma' : \Gamma', \Delta' \vdash C'}{
\deduce{\Sigma' : \Gamma' \vdash (\pi . B')\theta'}{\Pi_1} &
\deduce{\Sigma' : (\pi.B')\theta', \Delta' \vdash C'}{\Pi_2^{\theta',\pi,B'}}}
\end{equation*}
The proof of cut-elimination for $LG^\omega$ is based on induction
over the height of the right premise in a \cut, therefore this \cut\ can be
further reduced and eliminated. The essential properties we need to
complete the proof at this point are that $\Sigma' : \Gamma', \Delta'
\vdash C'$ is provable if and only if $\Sigma : \Gamma, \Delta \vdash
C$ is provable, and that both proofs have the same height in this
case. We formalize these in the lemma below.

\begin{defn}[Proof height]
The height of a derivation $\Pi$, denoted by $ht(\Pi)$, is $1$ if
$\Pi$ has no premise derivations and is the least
upper bound of $\{ht(\Pi_i)+1\}_{i\in\mathcal{I}}$ if $\Pi$ has the
premise derivations $\{\Pi_i\}_{i\in\mathcal{I}}$ where $\mathcal{I}$
is some index set. 
\end{defn}

\begin{lemma}[Raising]
\label{lem:raising}
Let $\Sigma : \Gamma \vdash C$ be a sequent, let $\vec{c}$ be a list of
nominal constants not in the support of $\Gamma$ or $C$, and let
$\Sigma' : \Gamma' \vdash C'$ be a version of $\Sigma : \Gamma \vdash C$
raised over $\vec{c}$. 
Then $\Sigma : \Gamma \vdash C$ has a proof of height $h$ if and only if
$\Sigma' : \Gamma' \vdash C'$ has a proof of height $h$. 
\end{lemma}

With this lemma in place, the following theorem and its corollary follow.

\begin{theorem}
The cut rule can be eliminated from \logic without affecting the
provability relation.
\end{theorem}

\begin{cor}
The logic \logic is consistent, \ie, it is not the case that both
$A$ and $A \supset \bot$ are provable.
\end{cor}

Cut-elimination is also useful in designing theorem provers 
and its counterpart, cut-admissibility, allows one to reason richly
about the properties of such proof procedures. 



\Section{Examples}
\label{sec:examples}

We will often suppress the outermost universal quantifiers in
displayed definitions and will assume that capital letters denote 
implicitly universally quantified variables.


\paragraph{Freshness} In Section~\ref{sec:logic} we showed how the
property of freshness could be defined in \logic by the definitional
clause
\begin{equation*}
\forall E.(\nabla x. \fresh x E) \triangleq \top.
\end{equation*}
This clause ensures that the atomic judgment $(\fresh X E)$ holds if
and only if $X$ is a nominal constant which does not appear anywhere
in the term $E$. To see the simplicity and directness of this
definition, consider how we might define freshness in a system like
$LG^\omega$ which allows for definitions only of atomic judgments. In
this situation, we 
will have to verify that $X$ is a nominal constant by ruling out the
possibility that it is a term of one of the other permitted
forms. Then, checking that $X$ does not appear in $E$ will require
an explicit walking over the structure of $E$. In short, such a
definition would have to have the specific structure of terms coded
into it and would also use (a mild form of) negative judgments.

\begin{figure}[t]
\begin{align*}
\member B L &\triangleq \exists n . \nat n \land \element n B L \\
\element z B (B::L) &\triangleq \top \\
\element {(s\; N)} B (C::L) &\triangleq \element N B L
\end{align*}
\vspace{-0.75cm}
\caption{List membership}
\label{fig:member}
\end{figure}

To illustrate how the definition in \logic can be used in a reasoning
task, consider proving the following lemma
\begin{equation*}
\forall x, e, \ell. (\fresh x \ell \land \member e \ell) \supset
\fresh x e
\end{equation*}
where {\sl member} is defined in Figure~\ref{fig:member}. This lemma
is useful in constructing arguments such as type uniqueness where one
must know that a list does not contain a typing judgment for a
particular variable. The proof of this lemma proceeds by induction on
the natural number $n$ quantified in the body of {\sl member}. The
base case and the inductive step eventually require showing the 
following:
\begin{align*}
\forall x, b, \ell.\ \fresh x (b :: \ell) \supset \fresh x b \\
\forall x, b, \ell.\ \fresh x (b :: \ell) \supset \fresh x \ell
\end{align*}
We shall consider the proof of only the first statement; the proof of
the second has a similar structure.

The first statement follows if we can prove the sequent
\begin{equation*}
x, b, \ell : \fresh x (b :: \ell) \vdash \fresh x b.
\end{equation*}
Consider how $\defL$ acts on the hypothesis $(\fresh x (b :: \ell))$ in
this sequent. First the clause for {\sl fresh} is raised over the
support of the hypothesis, but this is empty so raising has no effect.
Second, the sequent is raised over some new nominal constant $c$
corresponding to the $\nabla$ in the head of the definition for {\sl
  fresh}. The last step is to consider all permutations $\pi$ of the
set $\{c\}$ and all solutions $\theta$ of 
\begin{equation*}
(\pi. \fresh c e)\theta =
(\fresh {(x'\; c)} ((b'\; c) :: (\ell'\; c)))\theta.
\end{equation*}
There is, in fact, a most general unifier here:
\begin{tabbing}
\qquad\=$\theta = [$\=\kill
\>$\theta = [x' \to (\lambda x. x), b' \to (\lambda x. b''), $\\
\>\>$\ell' \to (\lambda x. \ell''), e \to (b'' :: \ell'')]$.
\end{tabbing}
The resulting sequent is
\begin{equation*}
b'', \ell'' : \top \vdash \fresh c b''
\end{equation*}

The next step in this proof is to apply $\defR$ to the conclusion. To
do this we first raise the clause for {\sl fresh} over the support of
the conclusion which is $\{c\}$. Then we raise the sequent over a new
nominal constant $c'$ corresponding to the $\nabla$ in the head of the
definition. Finally we need to find a permutation $\pi$ of $\{c,c'\}$
and a solution $\theta$ to $(\pi.\fresh {c'} (e'\; c))\theta = \fresh
c (b'''\; c')$. Here we find the permutation which swaps $c$ and $c'$
and the solution $\theta$ which unifies $e'$ and $b'''$. The resulting
sequent is then
\begin{equation*}
b''', \ell''' : \top \vdash \top
\end{equation*}
which is trivially provable.

\paragraph{Typing contexts} 

We now illustrate an approach to animating and reasoning about
the static and dynamic semantics of programming languages. The first
step in this approach is that of encoding these two kinds of semantics
using the (second-order fragment of the) logic of
hereditary Harrop formulas.
Specifications provided through these formulas have a natural
executable interpretation based on the logic 
programming paradigm \cite{miller91apal}. The interesting part from
the perspective of this paper is that we can encode provability of
this subset of hereditary Harrop formulas as a definition in
\logic. This definition, then, becomes the bridge for reasoning about
the (executable) specifications.

\begin{figure}[t]
\begin{align*}
\seq N L \langle A \rangle &\triangleq \member A L \\
\seq {(s\; N)} L (B \land C) &\triangleq \seq N L B \land \seq N L C \\
\seq {(s\; N)} L (A \supset B) &\triangleq \seq N {(A
  :: L)} B \\
\seq {(s\; N)} L (\forall B) &\triangleq \nabla x. \seq N L (B\; x) \\
\seq {(s\; N)} L \langle A\rangle  &\triangleq \exists b. \prog A b
\land \seq N L b
\end{align*}
\vspace{-0.75cm}
\caption{Second-order hereditary Harrop logic in \logic}
\label{fig:seq}
\end{figure}

To develop these ideas in more detail, we encode provability in the
second-order hereditary Harrop logic as a three-place 
definition $(\seq N L G)$ where $L$ denotes the context of hypothetical
(assumed) atomic formulas and $G$ denotes the goal formula
\cite{mcdowell02tocl,miller05tocl}.  The 
argument $N$ corresponds to the height of the proof tree and is used
for inductive arguments; we write this argument as a subscript to
downplay its significance. The definition of {\sl seq} is presented in
Figure \ref{fig:seq}.  The constructor $\langle \cdot \rangle$ is used
to inject atomic formulas into formulas; as such, it serves as a
device for isolating atomic formulas.  The object level universal
quantifier is reflected into a meta level generic (\ie, $\nabla$)
quantifier in the definition of {\em seq}; this treatment turns out to
capture the computational semantics of the universal quantifier rather
precisely. 
Backchaining is realized by the last clause of {\sl seq}. In giving
meaning to this clause, we expect that the specification of
interest in a particular situation (\ie, the {\em logic program} that we
want to reason about) has been encoded through the definition of
{\sl prog}. In particular, a logic program clause of 
the form $\forall \bar x.((G\ \bar x) \supset \tup{A\ \bar x})$ would
result, in the reasoning context, in the addition of 
a definitional clause $\forall \bar x.\prog{(A\ \bar x)}{(G\ \bar x)}
\triangleq \top$ that can be used by the {\sl seq} predicate.
To simplify notation, we write $L \tridot P$
for $\exists n . (\nat n \land \seq n L P)$. When $L$ is $nil$ we write
just $\,\tridot P$.

\begin{figure}
\begin{center}
\begin{tabular}{c}
$\forall m, n, t, u[\of m (\arr u t) \land
    \of n u \; \supset \; \of{(\app m n)} t]$\\[6pt] 
$\forall r, t, u[\forall x[\of x t  \supset 
    \of{(r \; x)}{u}] \supset \of{(\tabs t r)}{(\arr t u)}]$
\end{tabular}
\end{center}
\vspace{-0.25cm}
\caption{Simple typing of $\lambda$-terms} 
\label{fig:typing}
\end{figure}

An example of a specification that we may wish to reason about is that
of the typing rules for the simply typed $\lambda$-calculus. These
rules can be encoded using hereditary Harrop formulas as shown in Figure
\ref{fig:typing} that, in turn, would be reflected into definitional
clauses for {\sl prog} as described above. In these formulas, {\sl
  app} and {\sl abs} are  the usual constructors for application and
abstraction in the untyped $\lambda$-calculus. Note that no explicit
context of  typing assumptions is used in these rules: rather the
hypothetical judgment of hereditary Harrop formulas is used to keep
track of such assumptions. This context is made explicit only when
reasoning about this specification via the {\sl seq} definition.

Consider demonstrating the type uniqueness property for the simply typed
$\lambda$-calculus using the {\sl seq} encoding. We can do this by
showing that the formula
\begin{equation*}
\forall m, t, s. (\,\tridot \langle \of m t \rangle
\land \,\tridot \langle \of m s \rangle) \supset t = s,
\end{equation*}
is a theorem: here, the binary predicate $=$ is defined by the single
clause $\forall x.\ x = x \triangleq 
\top$. We can prove this formula using an induction on natural numbers
but, to do this, we must generalize it to account for the
fact that the rule for typing {\sl abs} that allows us to descend
under abstractions enhances the atomic formulas assumed by {\sl seq}. A
suitably generalized form of the statement, then, is
\begin{equation*}
\forall \ell, m, t, s. (\cntx \ell \land \ell \tridot \langle \of m
t \rangle \land \ell \tridot \langle \of m s \rangle) \supset t = s.
\end{equation*}
Now, this formula is provable only if the definition of $\cntx$ ensures
that if $\cntx \ell$ holds then $\ell$ is of the form \[(\of
{c_1} T_1::\ldots::\of {c_n} T_n::nil),\] where $c_1 \ldots c_n$ are distinct 
nominal constants. The challenge then, is in providing a definition of
{\sl cntx} which accurately describes this requirement. In particular, the
definition must ensure that the first arguments to {\sl of} in the
elements of this list are nominal constants and not some other piece
of syntax, and it must also ensure that each such constant is distinct
from all others. 

\begin{figure}[t]
\begin{align*}
& \cntx nil \triangleq \top \\
& \cntx (\of X A :: L) \triangleq (\forall M, N. X = \app
M N \supset \bot) \land\null \\
&\hspace{3.16cm} (\forall M, B . X = \tabs B M
\supset \bot) \land\null \\
&\hspace{3.16cm} (\forall B. \member {(\of X B)} L \supset
\bot) \land \null\\
&\hspace{3.16cm} \cntx L
\end{align*}
\vspace{-0.75cm}
\caption{{\sl cntx} in $LG^\omega$}
\label{fig:cntx-lg}
\end{figure}

\begin{figure}[t]
\begin{align*}
& \cntx nil \triangleq \top \\
(\nabla x. &\cntx (\of x A :: L)) \triangleq \cntx\; L
\end{align*}
\vspace{-0.75cm}
\caption{{\sl cntx} in \logic}
\label{fig:cntx-g}
\end{figure}

In $LG^\omega$, {\sl cntx} can be defined by explicitly restricting
each element of the context as shown in Figure \ref{fig:cntx-lg}.
This definition checks that the first argument to {\sl
of} is a nominal constant by explicitly ruling out all other
possibilities for it.
Then, to ensure distinctness of arguments, the rest of the list is
traversed using {\sl member}.
This definition is evidently complex and the complexity carries over
also into the process of reasoning based on it.

In \logic we can give a direct and concise definition of {\sl cntx}
using $\nabla$ quantification in the head of a definition as is done
in Figure \ref{fig:cntx-g}. The occurrence of the $\nabla$-bound
variable $x$ in 
the first argument of {\sl of} codifies the fact that type
assignments are only made for nominal constants. The uniqueness of such
nominal constants is enforced by the quantification structure of {\sl
  cntx}: the variable $L$ cannot contain any occurrences of $x$. With
this definition of {\sl cntx}, the generalized theorem of type
uniqueness is provable. Use of $\defL$ on the hypothesis of
$\cntx\ell$ will allow only the possibility of type assignments for
nominal constants, while use of $\defR$ will verify that the contexts
that are created in treating abstractions align with the requirements
imposed by the definition of {\sl cntx}. 

\paragraph{Arbitrarily cascading substitutions}


Reducibility arguments, such as Tait's proof of normalization for the
simply typed $\lambda$-calculus \cite{tait67jsl}, are based on
judgments over closed terms. During reasoning, however, one is often
working with open terms. To compensate, the closed term judgment is
extended to open terms by considering all possible closed
instantiations of the open terms. When reasoning with \logic, open
terms are denoted by terms with nominal constants representing free
variables. The general form of an open term is thus $M\; c_1\;
\cdots\; c_n$, and we want to consider all possible instantiations
$M\; V_1\; \cdots\; V_n$ where the $V_i$ are closed terms. This type of
arbitrary cascading substitutions is difficult to realize in reasoning
systems based on $\lambda$-tree syntax since $M$ would have an
arbitrary number of abstractions.

\begin{figure}[t]
\begin{align*}
& \subst z {nil} T T \triangleq \top \\
(\nabla x. &\subst {(s\; N)} {((x,V)::L)} {(T\; x)} S) \triangleq \\
&\hspace{4cm} \subst N L {(T\; V)} S
\end{align*}
\vspace{-0.75cm}
\caption{Arbitrary cascading substitutions}
\label{fig:subst}
\end{figure}

We can define arbitrary cascading substitutions in \logic using the
unique structure of definitions. In particular, we can define a
predicate which holds on a list of pairs $(c_i,V_i)$, a term
with the form $M\; c_1\; \cdots\; c_n$ and a term of the
form $M\; V_1\; \cdots\; V_n$. The idea is to iterate over the list of
pairs and for each pair $(c,V)$ use $\nabla$ in the head of a
definition to abstract $c$ out of the first term and then substitute
$V$ before continuing. This is the motivation for {\sl subst}
defined in Figure \ref{fig:subst}. Note that we have also added a
natural number argument to be used for inductive proofs.

Given the definition of {\sl subst} one may then show that arbitrary
cascading substitutions have many of the same properties as normal
higher-order substitutions. For instance, in the domain of the untyped
$\lambda$-calculus, we can show that {\sl subst} acts compositionally via
the following lemmas.
\begin{multline*}
\forall n, \ell, t, r, s. (\nat n
\land \subst n \ell {(\app t r)} s) \supset \\
\exists u, v. s = \app u v \land 
\subst n \ell t u \land \subst n \ell r v
\end{multline*}
\vspace{-1cm}
\begin{multline*}
\forall n, \ell, t, r. (\nat n
\land \subst n \ell {(\abs t)} r) \supset \\
\exists s. r = \abs s \land \nabla z. \subst n \ell {(t\; z)} (s\; z)
\end{multline*}
Both of these lemmas have straightforward proofs: induct on $n$, use
$\defL$ on the assumption of {\sl subst}, apply the inductive
hypothesis and use $\defR$ to complete the proof.



\Section{Related work}
\label{sec:related}

Mechanized reasoning about structural operational
se\-man\-tic-style specifications of formal systems has received 
the attention of other researchers.
Recent impetus for this kind of reasoning has been provided by a desire
for computer verified proofs in the realm of programming language
theory \cite{aydemir05tphols}.  One line of research focuses on
developing  proofs within the
framework provided by an existing and well-developed interactive theorem
prover such as Coq \cite{bertot04book} and Isabelle/HOL
\cite{nipkow02book}.
Many of the contexts in which machine authenticated
reasoning of this kind is needed deal with objects involving
binding. Several previous attempts have been characterized by the use
of algebraic datatypes, enhanced perhaps by a de Bruijn-like
representation of bound variables, in the encoding of binding
constructs. While some success has been achieved using this approach
to object representation
\cite{hirschkoff97tphol,leroy07tr,vaninwegen96phd},
it has also been noted that the real reasoning task is often  
overwhelmed under such an approach by the proofs of mundane binding
and substitution oriented lemmas. 

The more natural and more promising approaches to
the kind of reasoning of interest are the ones that provide special
logic based 
treatments of binding such as is manifest in $\lambda$-tree
syntax. We discuss the main lines of research under this rubric
below. 

\parag{Nominal logic based reasoning}
Nominal logic extends first-order syntax with primitives for treating
variable names in such a way that $\alpha$-equivalence classes
are recognized \cite{pitts03ic}. This considerably simplifies the
treatment of binding in specifications. In contrast to the approach
underlying our work, no separate meta-logic has as yet been developed
for reasoning about nominal logic descriptions. Reasoning about
specifications written in this logic is instead realized by 
axiomatizing the primitives of the logic in a rich system such as
Coq or Isabelle/HOL \cite{aydemir06lfmtp,urban05cade}. This approach has
proved successful for many applications.

Aside from the absence of a meta-logic, the most prominent difference
between the nominal logic based approach and 
our work is that we use $\lambda$-tree syntax and thus obtain a
comprehensive treatment of both $\alpha$-equivalence {\it and}
substitution within the logic. 
The nominal logic approach does not provide any direct support for
substitution, and instead requires substitution to be defined on a
case-by-case basis. In reasoning, this means that various substitution
lemmas need to be proved for each syntactic class over which substitution
is defined. Another difference worth noting is that we can derive
freshness as a consequence of the nesting of quantifiers in an explicit
definition of the {\sl fresh} predicate, whereas nominal logic
approaches either take freshness as primitive or define it in terms of
set membership. 

\parag{Two-levels of logic}
McDowell \& Miller \cite{mcdowell97phd,mcdowell97lics,mcdowell02tocl}
explored using a {\em two-level approach} to reasoning about, for
example, the operational semantics and the typing of small programming
languages.  Both levels of logic shared the same $\lambda$-tree
approach to the treatment of (object-level and meta-level) binding:
the object-logic was a simple second-order intuitionistic logic and
the meta-logic was called $\FOLDN$.  While $\FOLDN$ contained
inference rules for definitions, it lacked the $\nabla$-quantifier.  As a result,
the {\sl seq} predicate could not be specified in
the same direct fashion as it is in Figure~\ref{fig:seq}.

As we illustrated briefly in Section~\ref{sec:examples}, replacing
$\FOLDN$ with \logic strengthens the expressiveness of the meta-logic
by allowing more declarative approaches to the specification of
invariants for (object-level) contexts.  As a result, many of the
theorems that have been proved in $\FOLDN$ \cite{mcdowell02tocl} can
be given much more understandable proofs in \logic.

\parag{Twelf}
Pfenning and Schr\"umann \cite{schurmann98cade} also 
describe a two-level approach in which LF terms and types are
used at the object-level and the logic ${\cal M}_2$ is used at
the meta-level. Schr\"umann's PhD thesis \cite{schurmann00phd}
further extended that meta-logic to one called ${\cal M}_2^{+}$.
This framework is realized in Twelf \cite{pfenning99cade}, which
also provides a related style of meta-reasoning based on mode,
coverage, and termination checking over higher-order judgments in LF.
Their approach also makes use of $\lambda$-tree syntax at both the
object and meta-levels and goes beyond our proposal here in that they
handle the complexities of dependent types and proof objects
\cite{harper93jacm}. On the other hand, the kinds of meta-level
theorems they can prove are different from
what is available in \logic.  For example, implication and negation
are not present in ${\cal M}_2^{+}$ and cannot be encoded in
higher-order LF judgments: hence, properties
such as bisimulation for CCS or the $\pi$-calculus are not provable.

A key component in ${\cal M}_2^{+}$ and in the higher-order LF
judgment approach to meta-reasoning is the ability to specify
invariants related to the structure of meta-logical contexts. These
invariants are called {\it regular worlds} and their analogue in our
system is judgments such as {\sl cntx} which explicitly describe the
structure of contexts. While the approach to proving properties in
Twelf is powerful and convenient for many applications, one might
prefer defining explicit invariants, such as {\sl cntx}, over the use
of regular worlds, since this allows describing more general judgments
over contexts, such as in the example of arbitrary cascading
substitutions where the {\sl subst} predicate actively manipulates the
context of a term.


\parag{Implementation}
The first author has implemented a significant portion of \logic in a recently
released system called Abella \cite{gacek08abella}.
This system provides an interactive tactics-based interface to proof
construction.
The primary focus of Abella is on reasoning about object-level
specifications written in hereditary Harrop formulas: provability in
that logic is provided by a definition similar to that of {\sl seq} in
Figure~\ref{fig:seq}.
Through this approach, Abella is able to take advantage of meta-level
properties of the logic of hereditary Harrop formulas (\eg, cut and
instantiation properties) while never having to reason outside of
\logic.

Abella has been used in many applications, including all the examples
mentioned in this paper.
First-order results include reasoning on structures such as natural
numbers and lists.
Taking advantage of $\lambda$-tree syntax, application domains such as
the simply typed $\lambda$-calculus are directly accessible.
Particular results include equivalence of big-step and small-step
evaluation, preservation of typing for both forms of evaluation, and
determinacy for both forms of evaluation.
More advanced results which make use of generic judgments for
describing contexts include type uniqueness, disjoint partitioning
of $\lambda$-terms into normal and non-normal form, and the
Church-Rosser theorem.
Larger applications include challenges 1a and 2a of the POPLmark
challenge \cite{aydemir05tphols}, a task which involves reasoning
about the contexts of subtyping judgments for $F_{<:}$, a
$\lambda$-calculus with bounded subtype polymorphism.
Finally, we have formalized a proof of normalization for the
simply-typed $\lambda$-calculus based on Tait's reducibility argument
\cite{tait67jsl}. This last example uses the formalization of
arbitrarily cascading substitutions described Section
\ref{sec:examples}.

\Section{Future work}
\label{sec:future}

We are presently investigating the extension of \logic with a general
treatment of induction over definitions as in the closely
related logic Linc \cite{tiu04phd}. This extension would 
simplify many inductive arguments by
obviating explicit measures in
induction; thus, natural numbers encoding computation lengths
would not be needed in the definitions of the {\sl   element} and {\sl
  subst} predicates considered in  
Section~\ref{sec:examples} if we can induct directly on the unfolding of
their definitions. Another benefit of this approach to induction is that it
has a naturally dual rule for coinduction over coinductive
definitions. This rule has been found useful in Linc, for example, in
proving properties of systems such as the $\pi$-calculus.

\ignore{
The Linc logic in \cite{tiu04phd} allows for a general way to take
recursive definitions and yield inductive inference rules for them:
thus, Linc does not require providing defined predicates with explicit
natural number measures (such as the subscript arguments to {\sl seq}
and {\sl subst}) in order to perform induction: we hope to extend
\logic to allow similar inductive inference rules and to provide also
Linc's notions of coinductive inference rules.
}

At a practical level, we are continuing to develop Abella as a theorem
proving system and to explore its use in complex reasoning tasks. We
expect to use Abella to provide more elegant proofs of the many
meta-logical theorems found in \cite{mcdowell02tocl}, which include
cut-elimination theorems, type preservation, and determinacy of typing
and evaluation. Finally, if the previously mentioned work on
coinduction is completed, Abella can be used to explore the role of
generic definitions in a coinductive setting.




\Section{Acknowledgements}

We thank David Baelde and Alwen Tiu for valuable suggestions and
anonymous reviewers for comments on an earlier version of this paper.
This work has been supported by 
INRIA through the ``Equipes Associ{\'e}es'' Slimmer, by the NSF
Grants OISE-0553462 (IRES-REUSSI) and CCR-0429572, and by a grant from
Boston Scientific. Opinions, findings, and conclusions or 
recommendations expressed in this papers are those of the authors and
do not necessarily reflect the views of the National Science
Foundation.


\bibliographystyle{latex8}
\bibliography{../references/master}

\end{document}